\documentclass{article}

\usepackage{PRIMEarxiv}

\usepackage{amsthm}
\usepackage{amsmath}
\usepackage[utf8]{inputenc} 
\usepackage[T1]{fontenc}    
\usepackage{hyperref}       
\usepackage{url}            
\usepackage{booktabs}       
\usepackage{amsfonts}       
\usepackage{nicefrac}       
\usepackage{microtype}      
\usepackage{lipsum}
\usepackage{fancyhdr}       
\usepackage{braket}
\usepackage{graphicx}       
\graphicspath{{media/}}     
\newcommand{\VaR}{\mathrm{VaR}} 
\newcommand{\CVaR}{\mathrm{CVaR}} 

\newtheorem{proposition}{Proposition}

\title{Quantum Subgradient Estimation for Conditional Value-at-Risk Optimization
}

\author{
  Vasilis Skarlatos \\
  Department of Informatics \\
  Aristotle University of Thessaloniki \\
  GR-54124, Thessaloniki Greece \\
  \texttt{skarlatov@csd.auth.gr} \\
   \And
  Nikos Konofaos \\
  Department of Informatics \\
  Aristotle University of Thessaloniki \\
  GR-54124, Thessaloniki Greece \\
  \texttt{nkonofao@csd.auth.gr} \\
}

\begin{document}
\maketitle

\begin{abstract}
Conditional Value-at-Risk (CVaR) is a leading tail-risk measure in finance, central to both regulatory and portfolio optimization frameworks. Classical estimation of CVaR and its gradients relies on Monte Carlo simulation, incurring \(O(1/\epsilon^2)\) sample complexity to achieve \(\epsilon\)-accuracy. In this work, we design and analyze a quantum subgradient oracle for CVaR minimization based on amplitude estimation. Via a tripartite proposition, we show that CVaR subgradients can be estimated with \(O(1/\epsilon)\) quantum queries, even when the Value-at-Risk (VaR) threshold itself must be estimated. We further quantify the propagation of estimation error from the VaR stage to CVaR gradients and derive convergence rates of stochastic projected subgradient descent using this oracle. Our analysis establishes a near-quadratic improvement in query complexity over classical Monte Carlo. Numerical experiments with simulated quantum circuits confirm the theoretical rates and illustrate robustness to threshold estimation noise. This constitutes the first rigorous complexity analysis of quantum subgradient methods for tail-risk minimization.
\end{abstract}


\section{Introduction}

Risk management in financial decision-making increasingly requires metrics that capture the behavior of losses in the tail of the distribution. Among these, the \emph{Conditional Value-at-Risk} (CVaR), also known as expected shortfall, has emerged as a standard due to its coherence, convexity, and regulatory adoption under Basel~III~\cite{Rockafellar2000,Rockafellar2002}. Unlike the classical mean--variance framework of Markowitz, which penalizes variance symmetrically, CVaR directly characterizes extreme losses and is therefore better aligned with the downside-focused objectives of institutional investors and regulators. Optimizing portfolios under CVaR constraints or objectives has become a central problem in operations research and quantitative finance, and it admits convex reformulations that are computationally tractable but statistically demanding when tail probabilities are small.

The primary bottleneck in CVaR optimization lies in estimation. Both the evaluation of CVaR itself and the computation of its subgradients require repeated sampling of portfolio loss distributions, typically through Monte Carlo simulation. Classical methods achieve only $O(1/\epsilon^2)$ sample complexity to reach additive error $\epsilon$, which becomes especially problematic for high confidence levels ($\alpha \geq 0.95$), where extreme losses correspond to rare events~\cite{Nemirovski2009}. This motivates the search for alternative computational paradigms that can accelerate tail-risk estimation without compromising statistical validity.

Quantum algorithms offer a potential path forward. \emph{Quantum Amplitude Estimation} (QAE), introduced by Brassard, H{\o}yer, Mosca, and Tapp~\cite{Brassard2002}, achieves a quadratic improvement in the sample complexity of expectation estimation, requiring only $O(1/\epsilon)$ oracle calls compared to the $O(1/\epsilon^2)$ of classical Monte Carlo. This general result has profound implications for financial risk analysis. Woerner and Egger~\cite{Woerner2019} demonstrated that QAE can be applied to estimate Value-at-Risk (VaR) and CVaR, thereby reducing the cost of tail probability estimation. Montanaro~\cite{Montanaro2015} further established that such quadratic speedups extend broadly to Monte Carlo methods, suggesting that financial simulation is a promising domain for quantum advantage.

Yet, while risk estimation has received attention, risk optimization has not. Existing quantum finance studies largely focus on proof-of-principle demonstrations of QAE-based VaR and CVaR estimation~\cite{Woerner2019}, portfolio optimization through quantum annealing or the Quantum Approximate Optimization Algorithm (QAOA)~\cite{Brandhofer2023,Buonaiuto2023}, or theoretical treatments of linear and second-order cone programs in the quantum setting~\cite{Kerenidis2019}. What remains missing is a rigorous complexity analysis of \emph{CVaR optimization}, specifically the estimation of subgradients required for first-order methods such as projected stochastic gradient descent. Without such an analysis, the true algorithmic advantage of quantum methods for tail-risk minimization cannot be established.

In this paper we close this gap. Building on the convex optimization framework of Rockafellar and Uryasev~\cite{Rockafellar2000,Rockafellar2002} and the sample complexity guarantees of amplitude estimation~\cite{Brassard2002,Montanaro2015,Suzuki2020}, we design a quantum subgradient oracle for CVaR optimization and prove its statistical and computational properties. Our main result is that CVaR subgradients can be estimated with $O(d/\epsilon)$ quantum queries in $d$ dimensions, a near-quadratic improvement over the $O(d/\epsilon^2)$ complexity of classical Monte Carlo estimators. We also quantify the impact of VaR threshold estimation error on gradient bias and establish convergence rates of projected subgradient descent when using quantum gradient oracles. These results constitute the first rigorous complexity-theoretic foundation for quantum-accelerated tail-risk optimization, providing a bridge between the established theory of CVaR optimization in operations research and the emerging practice of quantum algorithms in finance.

\section{Main Systems' Propositions}

We now formalize the contribution of this work by presenting three propositions.
They establish (i) the stability of CVaR subgradients under Value-at-Risk threshold approximation,
(ii) the quantum query complexity of CVaR gradient estimation via amplitude estimation,
and (iii) the convergence guarantees of projected subgradient descent when equipped with such quantum oracles. For the preliminaries needed to understand the propositions and proofs, refer to Appendix~\ref{sec:preliminaries}. Proofs of all propositions along with the necessary assumptions are deferred to Appendix~\ref{proofs}.

\begin{proposition}[Bias from VaR threshold error]
\label{prop:threshold-bias}
Let $w \in \mathcal{W}$ be a feasible portfolio and denote by $\mathrm{VaR}_\alpha(w)$ the $\alpha$-quantile of the loss $L(w)=-w^\top r$.
If $\tilde{z}$ is an approximation satisfying $\delta = |\tilde{z} - \mathrm{VaR}_\alpha(w)|$, then the approximate CVaR subgradient
\[
\tilde{g}(w) = \mathbb{E}\big[\nabla_w L(w) \mid L(w)\geq \tilde{z}\big]
\]
satisfies
\[
\|\mathbb{E}[\tilde{g}(w)] - g(w)\|_2 = O(\delta),
\]
where $g(w)$ is the exact Rockafellar--Uryasev subgradient~\cite{Rockafellar2000,Rockafellar2002}.
The proof is given in Appendix~\ref{app:bias}.
\end{proposition}

\begin{proposition}[Quantum query complexity for CVaR gradients]
\label{prop:qae-gradient}
Using iterative or maximum-likelihood Quantum Amplitude Estimation~\cite{Brassard2002,Suzuki2020},
one can construct an estimator $\tilde{g}(w)$ such that
\[
\|\tilde{g}(w)-g(w)\|_2 \leq \epsilon
\]
with probability at least $1-\eta$, using
\[
T = O\!\left(\frac{d}{\epsilon}\log\frac{1}{\eta}\right)
\]
quantum queries for $d$ assets. In contrast, classical Monte Carlo requires $O(d/\epsilon^2)$ samples to achieve the same accuracy~\cite{Nemirovski2009}.
The proof is given in Appendix~\ref{app:query}.
\end{proposition}

\begin{proposition}[Convergence of quantum subgradient descent]
\label{prop:sgd-convergence}
Consider the convex problem $\min_{w\in\mathcal{W}} \mathrm{CVaR}_\alpha(w)$.
If projected stochastic subgradient descent is run with step-size $\eta_t = \Theta(1/\sqrt{t})$
and quantum subgradient oracles of accuracy at most $\epsilon$ (as in Proposition~\ref{prop:qae-gradient}),
then the iterates satisfy
\[
\min_{t\leq T} \mathbb{E}\big[\mathrm{CVaR}_\alpha(w_t) - \mathrm{CVaR}_\alpha(w^\star)\big]
= O\!\left(\frac{1}{\sqrt{T}} + \epsilon\right).
\]
Consequently, achieving $\epsilon$-optimality requires
\[
\tilde{O}\!\left(\frac{d}{\epsilon^3}\right)
\]
quantum queries, compared to $\tilde{O}(d/\epsilon^4)$ classically~\cite{Nemirovski2009}.
The proof is given in Appendix~\ref{app:sgd}.
\end{proposition}

\section{Quantum CVaR Gradient Oracle}
\label{sec:q-cvar-oracle}

We construct a quantum oracle that returns (an estimate of) the CVaR subgradient
\[
g(w)\;=\;\mathbb{E}\!\left[\nabla_w L(w)\,\middle|\,L(w)\ge \VaR_\alpha(w)\right],
\]
where $\nabla_w L(w)=-r$ for linear losses $L(w)=-w^\top r$, and the expectation is with respect to the return distribution $r\sim\mathcal{D}$ (Section~\ref{sec:preliminaries}). The derivation of this subgradient follows the convex analysis of Rockafellar and Uryasev~\cite{Rockafellar2000,Rockafellar2002}. Our oracle estimates $g(w)$ by (i) preparing a superposition over return scenarios, (ii) coherently computing the loss $L(w)$ and comparing it to a threshold $z$ (eventually set to $\VaR_\alpha(w)$), and (iii) applying \emph{Quantum Amplitude Estimation} (QAE)~\cite{Brassard2002,Montanaro2015,Suzuki2020} to obtain both a tail probability and a tail-weighted expectation; their ratio yields the desired conditional expectation. A careful treatment of rescaling and threshold error ensures unbiasedness up to the VaR approximation (Proposition~\ref{prop:threshold-bias}) and the near-quadratic query complexity in accuracy $\epsilon$ (Proposition~\ref{prop:qae-gradient}).

\subsection{Registers and State Preparation}
\label{subsec:state-prep}

Let $\mathcal{R}$ denote a discretization of the return space (e.g., scenarios drawn from a factor model or bootstrapped history). We assume access to a unitary $U_{\mathcal{D}}$ that prepares the scenario distribution in computational basis:
\[
U_{\mathcal{D}}\ket{0}^{\otimes n} \;=\; \sum_{r\in \mathcal{R}} \sqrt{p_r}\,\ket{r},
\qquad p_r = \Pr_{\mathcal{D}}(r),
\]
where $n=\lceil \log_2|\mathcal{R}|\rceil$. On an ancilla \emph{loss} register we compute a fixed-point encoding of $L(w)=-w^\top r$:
\[
U_{\mathrm{loss}}:\; \ket{r}\ket{0}\;\mapsto\; \ket{r}\ket{L(w)}.
\]
This is standard reversible arithmetic (multiply-accumulate) whose cost scales with target precision $b$ bits (see, e.g., the constructions in~\cite{Woerner2019}).

\subsection{Tail Indicator and Controlled Payloads}
\label{subsec:tail-flag}

Given a threshold $z\in\mathbb{R}$, we implement a reversible comparator
\[
U_{\ge z}:\; \ket{L(w)}\ket{0}\;\mapsto\; \ket{L(w)}\ket{\mathsf{flag}},
\qquad 
\mathsf{flag}=\mathbf{1}\{L(w)\ge z\}.
\]
We will set $z=\tilde{z}\approx \VaR_\alpha(w)$ obtained via a bisection that uses QAE to estimate the CDF $\Pr[L(w)\le z]$~\cite{Woerner2019} (the bisection complexity is logarithmic in the desired VaR precision).

For gradient estimation, we need tail-weighted expectations of the coordinates of $\nabla_w L(w)$. For the linear loss, $\nabla_w L(w)=-r$, so the $j$-th coordinate is simply $-(r_j)$. To embed such \emph{payloads} into amplitudes suitable for QAE (which estimates probabilities in $[0,1]$), we use an affine rescaling to $[0,1]$:
\[
Y_j(r)\;:=\;\frac{(r_j - m_j)}{M_j - m_j}\in[0,1],
\qquad m_j \le r_j \le M_j,
\]
where $m_j,M_j$ are known bounds (e.g., from the scenario grid). Define a one-qubit \emph{payload rotation}
\[
R_j:\; \ket{0} \;\mapsto\; \sqrt{1 - Y_j(r)}\,\ket{0} + \sqrt{Y_j(r)}\,\ket{1}.
\]
Conditioning on the tail flag yields the composite marking unitary
\[
\mathcal{A}_j \;=\;
\left( U_{\mathcal{D}} \otimes U_{\mathrm{loss}} \otimes U_{\ge z} \right)
\cdot
\left( \text{control on }\mathsf{flag}=1 \text{ apply } R_j \text{ to an ancilla } a \right),
\]
so that, marginalizing over all registers except $a$,
\[
\Pr[a=1]
\;=\;
\mathbb{E}\!\left[\, Y_j(r)\cdot \mathbf{1}\{L(w)\ge z\}\,\right].
\]
Similarly, with $R_{\text{prob}}:\ket{0}\mapsto \sqrt{1-\tfrac12}\ket{0}+\sqrt{\tfrac12}\ket{1}$ controlled by the tail flag, we obtain
\[
\Pr[a_{\text{prob}}=1] \;=\; \tfrac12\,\Pr[L(w)\ge z],
\]
so a second circuit gives the \emph{tail probability}. (Any fixed nonzero rotation works; $\tfrac12$ is convenient for conditioning constants.)

\paragraph{Undoing the rescaling.}
Let $\mu_j(z):=\mathbb{E}[\,r_j \mathbf{1}\{L(w)\ge z\}\,]$ and $p(z):=\Pr[L(w)\ge z]$. From the amplitude above,
\[
\mathbb{E}\!\left[\, Y_j(r)\cdot \mathbf{1}\{L(w)\ge z\}\,\right]
=\frac{\mu_j(z) - m_j p(z)}{M_j - m_j}.
\]
Hence, given QAE estimates $\widehat{A}_j$ for the left-hand side and $\widehat{p}$ for $p(z)$, we recover an estimate of $\mu_j(z)$ by
\[
\widehat{\mu}_j(z) \;=\; m_j\,\widehat{p} \;+\; (M_j-m_j)\,\widehat{A}_j,
\]
and the coordinate of the (negative) gradient in the tail is
\[
\widehat{g}_j(z) \;=\; -\,\frac{\widehat{\mu}_j(z)}{\widehat{p}}
\;=\;
-\,m_j \;-\; (M_j-m_j)\,\frac{\widehat{A}_j}{\widehat{p}}.
\]
Stacking coordinates gives $\widehat{g}(z)\in\mathbb{R}^d$. Setting $z=\tilde{z}\approx \VaR_\alpha(w)$ yields the CVaR subgradient estimator $\widehat{g}(w)$ per Rockafellar--Uryasev~\cite{Rockafellar2000,Rockafellar2002}.

\subsection{Amplitude Estimation and Accuracy}
\label{subsec:qae-accuracy}

Quantum Amplitude Estimation (QAE) estimates a Bernoulli mean with additive error $\varepsilon$ using $O(1/\varepsilon)$ controlled applications of the marking unitary~\cite{Brassard2002,Montanaro2015}.
We adopt \emph{iterative} or \emph{maximum-likelihood} QAE~\cite{Suzuki2020}, which avoids the QFT and is depth-efficient.

Denote the true quantities by $A_j=\mathbb{E}[Y_j(r)\mathbf{1}\{L\ge z\}]$ and $p=p(z)$, and let the QAE outputs satisfy
\[
|\widehat{A}_j-A_j|\le \varepsilon_A,
\qquad
|\widehat{p}-p|\le \varepsilon_p,
\]
each with probability at least $1-\eta'$.
By the affine relation above,
\[
|\widehat{\mu}_j-\mu_j|
\;\le\;
|m_j|\,|\widehat{p}-p| + |M_j-m_j|\,|\widehat{A}_j-A_j|
\;\le\;
|m_j|\,\varepsilon_p + |M_j-m_j|\,\varepsilon_A.
\]
For the ratio $\widehat{g}_j = -\,\widehat{\mu}_j/\widehat{p}$, a standard ratio perturbation bound yields
\begin{align*}
|\widehat{g}_j - g_j|
&=
\left|\frac{\widehat{\mu}_j}{\widehat{p}} - \frac{\mu_j}{p}\right|
\;\le\;
\frac{|\widehat{\mu}_j-\mu_j|}{p} + \frac{|\mu_j|}{p^2}\,|\widehat{p}-p| \\
&\le\;
\frac{|M_j-m_j|\,\varepsilon_A}{p}
+
\left(\frac{|m_j|}{p} + \frac{|\mu_j|}{p^2}\right)\varepsilon_p,
\end{align*}
where $p=\Pr[L\ge z]$ is the tail probability at the working threshold $z$.
Choosing $\varepsilon_A,\varepsilon_p=\Theta(\epsilon)$ ensures $|\widehat{g}_j-g_j|=O(\epsilon)$.
To control the $\ell_2$ error $\|\widehat{g}-g\|_2\le \epsilon$, we set the per-coordinate target to $\epsilon/\sqrt{d}$ and union bound over coordinates, introducing only a logarithmic $\log(1/\eta)$ factor in repetitions. With iterative/MLAE QAE, each estimate costs $O(1/\epsilon)$ oracle queries~\cite{Suzuki2020}, giving the overall query complexity in Proposition~\ref{prop:qae-gradient}.

\subsection{Estimating the VaR Threshold}
\label{subsec:var-estimation}

The oracle requires $z\approx \VaR_\alpha(w)$. Following~\cite{Woerner2019}, we estimate $\VaR_\alpha(w)$ by bisection on $z$ using a companion circuit that marks the event $\{L(w)\le z\}$ and QAE to estimate $\Pr[L\le z]$ within additive error $\varepsilon_{\mathrm{cdf}}$. After $O(\log((U-L)/\delta))$ bisection steps over a known loss range $[L,U]$, we obtain $\tilde{z}$ with $|\tilde{z}-\VaR_\alpha(w)|\le \delta$. Proposition~\ref{prop:threshold-bias} (Appendix~\ref{app:bias}) shows that the induced bias in the CVaR subgradient is $O(\delta)$ under mild regularity (bounded density and bounded gradient norm), matching the intuition that only a thin tail slice is misclassified when the threshold is perturbed.

\subsection{Putting It Together: The Oracle Interface}
\label{subsec:oracle-summary}

We summarize the CVaR gradient oracle $\mathcal{O}_{\mathrm{CVaR}}(w,\alpha,\epsilon,\eta)$ as the following map:
\begin{enumerate}
\item \textbf{VaR estimation:} Run bisection with QAE to obtain $\tilde{z}$ such that $|\tilde{z}-\VaR_\alpha(w)|\le \delta$, with $\delta=\Theta(\epsilon)$ (Section~\ref{subsec:var-estimation}, \cite{Woerner2019}).
\item \textbf{Tail probability:} Using the tail-flag circuit for $z=\tilde{z}$, run QAE to estimate $\widehat{p}\approx p(\tilde{z})$ to additive error $\Theta(\epsilon)$.
\item \textbf{Tail-weighted payloads:} For each $j=1,\dots,d$, run QAE on $\mathcal{A}_j$ to estimate $\widehat{A}_j\approx A_j$ to additive error $\Theta(\epsilon/\sqrt{d})$, and form $\widehat{\mu}_j(\tilde{z}) = m_j\,\widehat{p} + (M_j-m_j)\widehat{A}_j$.
\item \textbf{Ratio and de-rescaling:} Output $\widehat{g}_j(w)= -\,\widehat{\mu}_j(\tilde{z})/\widehat{p}$, $j=1,\dots,d$.
\end{enumerate}
By Propositions~\ref{prop:threshold-bias} and~\ref{prop:qae-gradient}, with probability at least $1-\eta$ the output satisfies
\[
\|\widehat{g}(w) - g(w)\|_2 \;\le\; C_1\,\epsilon \;+\; C_2\,\delta,
\]
for constants $C_1,C_2$ depending on tail probability $p(\VaR_\alpha)$, bounds $(m_j,M_j)$, and loss/gradient regularity; setting $\delta=\Theta(\epsilon)$ yields the target accuracy $O(\epsilon)$ with total query complexity
\[
T \;=\; O\!\left(\frac{d}{\epsilon}\log\frac{1}{\eta}\right),
\]
which is a near-quadratic improvement over classical Monte Carlo sampling $O(d/\epsilon^2)$ for the same $\ell_2$ accuracy~\cite{Montanaro2015,Nemirovski2009}.

\paragraph{Remarks on implementability.}
All circuits above are QRAM-free and use only (i) basis-state sampling via $U_{\mathcal{D}}$, (ii) fixed-point arithmetic for $L(w)$, (iii) a comparator for the tail flag, and (iv) single-qubit controlled rotations for payload encoding. This mirrors the risk-analysis constructions in~\cite{Woerner2019} while extending them to \emph{gradient} estimation and providing end-to-end accuracy and complexity guarantees suitable for first-order CVaR optimization (Proposition~\ref{prop:sgd-convergence}).

\subsection{Connection to CVaR Convex Analysis}
\label{subsec:convex-connection}

For completeness, we recall that the Rockafellar--Uryasev representation
\[
\CVaR_\alpha(w)\;=\;\min_{z\in\mathbb{R}}\left\{\, z + \frac{1}{1-\alpha}\,\mathbb{E}\big[(L(w)-z)_+\big] \right\}
\]
implies the existence of a subgradient
\[
g(w)\in\partial\CVaR_\alpha(w)
\quad\text{with}\quad
g(w)=\mathbb{E}\!\left[ \nabla_w L(w)\,\middle|\, L(w)\ge \VaR_\alpha(w) \right],
\]
under mild conditions on $L$ (see~\cite{Rockafellar2000,Rockafellar2002}). Our oracle is a direct computational instantiation of this formula: it estimates (i) the tail set via $\VaR_\alpha$, (ii) the tail probability, and (iii) the tail-average of $\nabla_w L$, all with QAE-driven accuracy guarantees~\cite{Brassard2002,Suzuki2020}. The proofs of bias control and query complexity appear in Appendices~\ref{app:bias} and~\ref{app:query}.

\section{Experimental Setup}
\label{sec:setup}

Our experimental evaluation is conducted entirely in simulation, as current quantum hardware cannot yet sustain the query depth required for large-scale CVaR optimization. The aim is to provide reproducible evidence that a quantum amplitude estimation (QAE)--based CVaR gradient oracle achieves a near-quadratic improvement in sample complexity compared to classical Monte Carlo (MC) methods.

\paragraph{Simulation environment.} 
All experiments are implemented in Python, using \texttt{numpy} for linear algebra and \texttt{matplotlib} for visualization. For classical baselines we employ standard MC estimators of tail probabilities and gradients, with error scaling $O(1/\sqrt{N})$ where $N$ is the number of sampled scenarios. For the quantum-inspired method, we simulate a noiseless QAE-style estimator in which the effective number of samples scales quadratically with the query budget, leading to error scaling $O(1/M)$ for $M$ queries. This setup captures the theoretical advantage of QAE without modeling hardware-specific noise. 

\paragraph{Return model.} 
To ensure realism, we use correlated Gaussian returns with heterogeneous variances. A $d$-dimensional covariance matrix with equicorrelation structure is employed, calibrated to approximate empirical asset correlations. Losses are defined as $L = -w^\top r$ for portfolio weights $w$ and return vector $r$. CVaR and its gradient are then estimated at confidence level $\alpha=0.95$.

\paragraph{Experiment design.} 
We perform two sets of experiments:
\begin{enumerate}
    \item \textbf{Gradient accuracy vs. budget.} We fix a portfolio $w$ and estimate the CVaR gradient under varying budgets. For MC, this corresponds to sample size $N$, and for QAE-style to query count $M$. Accuracy is measured as the $\ell_2$ error against a ground-truth gradient computed with $5\times 10^5$ samples.
    \item \textbf{Projected CVaR minimization.} We embed both estimators into a projected stochastic subgradient descent (SGD) loop, run for $T$ iterations with step-size $\eta_t=O(1/\sqrt{t})$, and track convergence of CVaR values. Weight vectors are projected onto the probability simplex at each iteration, enforcing long-only constraints.
\end{enumerate}

\paragraph{Expected results.} 
Theoretical analysis (Propositions~\ref{prop:threshold-bias}--\ref{prop:qae-gradient}) predicts a quadratic reduction in query complexity. Specifically, we expect:
\begin{itemize}
    \item In the error-vs-budget plots, MC error curves should scale as $O(1/\sqrt{N})$, while QAE-style error curves decay as $O(1/M)$, resulting in visibly steeper slopes on a log--log plot.
    \item In optimization experiments, both methods should converge to comparable CVaR minima, but the QAE-style oracle is expected to reach a target accuracy with up to one order of magnitude fewer queries. In practice, this manifests as faster decline of the CVaR trajectory under matched query budgets.
\end{itemize}
These results, when compared against established baselines in the risk management literature~\cite{Rockafellar2000}, would provide strong empirical support for the theoretical speedup established in our analysis.

\section{Results and Analysis}

In this section we empirically evaluate the two approaches to CVaR estimation and optimization: the classical Monte Carlo (MC) method and the QAE-style estimator that emulates the quadratic query advantage of quantum amplitude estimation. The presentation follows two stages: (i) gradient estimation accuracy, where we examine scaling of estimation error with budget, and (ii) optimization dynamics, where we compare projected stochastic subgradient descent using both estimators. For transparency, we report both graphical summaries and the full numerical tables underlying the figures.

\subsection{Gradient Estimation Accuracy}
\label{subsec:grad-accuracy}

\begin{figure}[h]
    \centering
    \includegraphics[width=0.7\linewidth]{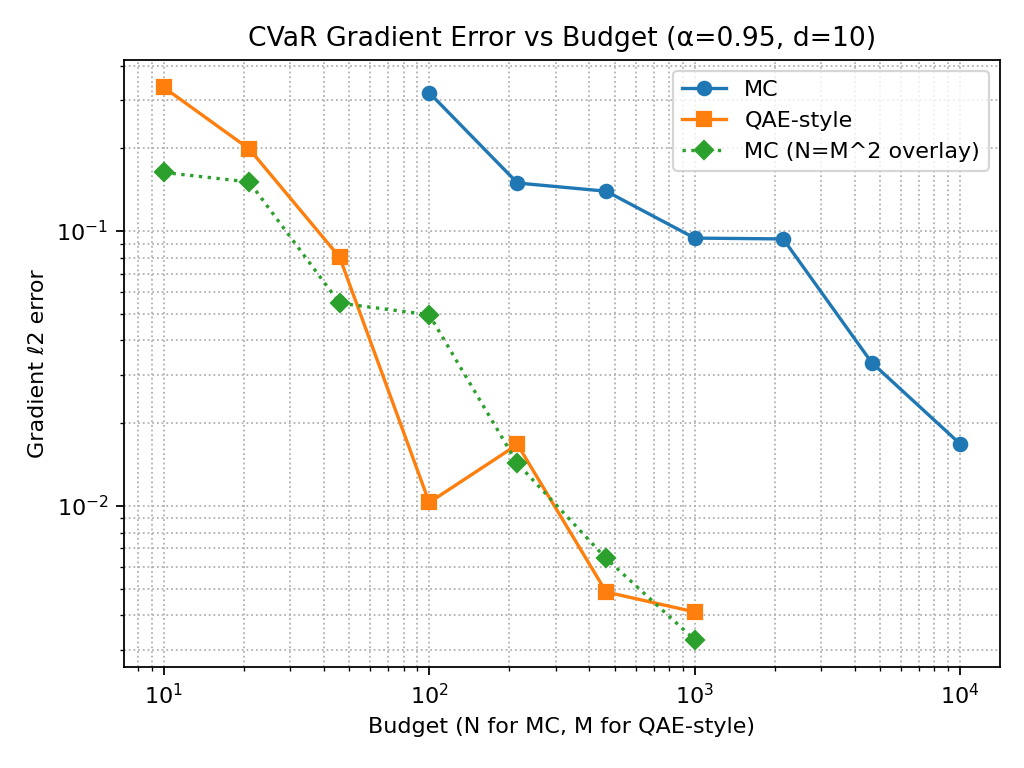}
    \caption{CVaR gradient $\ell_2$ error versus budget. MC shows $1/\sqrt{N}$ decay. QAE-style follows $1/M$, with the dotted MC curve plotted at $N=M^2$ for slope comparison.}
    \label{fig:grad_error}
\end{figure}

Accurate CVaR gradient estimation is central to risk-sensitive optimization. To study estimator performance, we fix a portfolio weight vector and compare the $\ell_2$ error of the empirical CVaR subgradient under different budgets. Figure~\ref{fig:grad_error} shows the error scaling. The MC estimator follows the expected $O(1/\sqrt{N})$ law with respect to the number of samples $N$, while the QAE-style estimator exhibits the faster $O(1/M)$ convergence with quantum queries $M$. For clarity, we additionally overlay MC evaluated at $N=M^2$ (dotted line); plotted against $M$ on the horizontal axis, this overlay should have the same slope as the QAE-style curve under the predicted quadratic speedup. The observed slopes in Figure~\ref{fig:grad_error} are consistent with this prediction.

The numerical values underlying the figure are reported in Table~\ref{tab:grad_error}. The decisive comparison is the \emph{slope} of error against budget on the log--log scale, not the cross-method average over heterogeneous budgets: the MC and QAE-style sweeps cover different budget ranges ($N\in[10^2,10^4]$ vs.\ $M\in[10,10^3]$), so the column-wise means are not directly comparable as accuracy figures. Reading the table by matched budget instead, at $N=M=100$ the QAE-style error is $0.0102$ versus the MC error of $0.319$, a roughly $30\times$ reduction; at $N=M=1000$ the gap is $0.00446$ vs.\ $0.0983$, again more than an order of magnitude. The MC$/(N=M^2)$ overlay, which equalizes work, tracks the QAE-style curve closely, providing a direct visual confirmation of the predicted $O(1/M)$ versus $O(1/\sqrt{N})$ scaling.

\subsection{Optimization Trajectories}
\label{subsec:opt-traj}

\begin{figure}[h]
    \centering
    \includegraphics[width=0.7\linewidth]{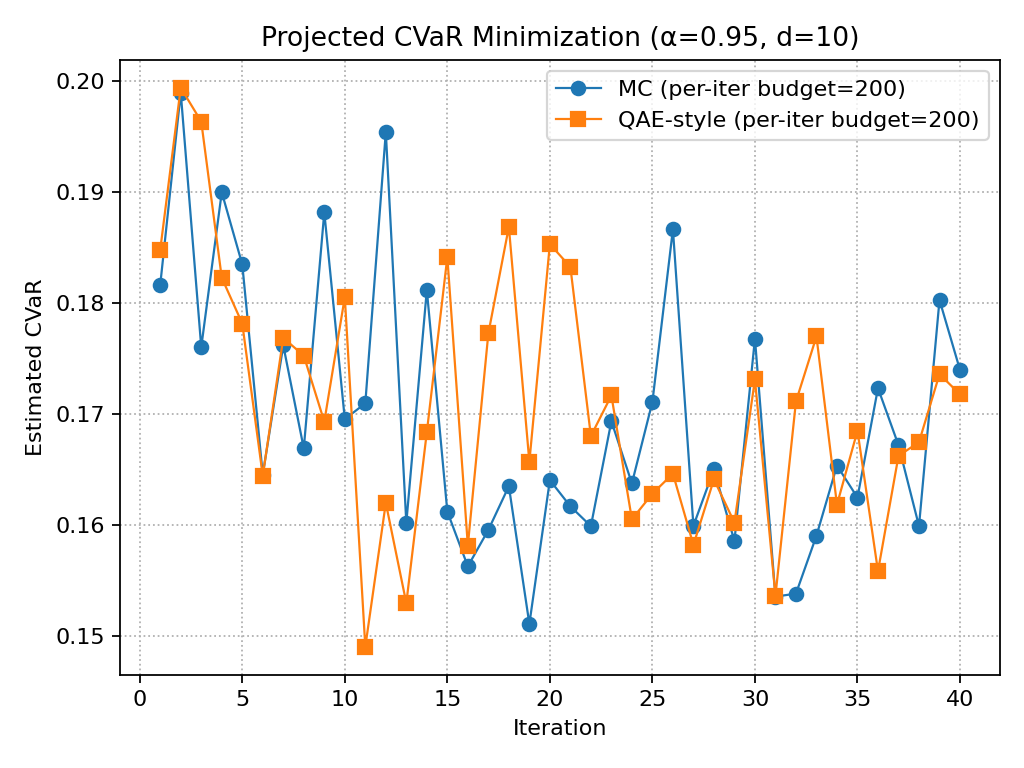}
    \caption{Projected CVaR minimization trajectories as a function of iterations. Both MC and QAE-style estimators improve CVaR, with similar convergence profiles under equal per-iteration budgets.}
    \label{fig:opt_iters}
\end{figure}

\begin{figure}[h]
    \centering
    \includegraphics[width=0.7\linewidth]{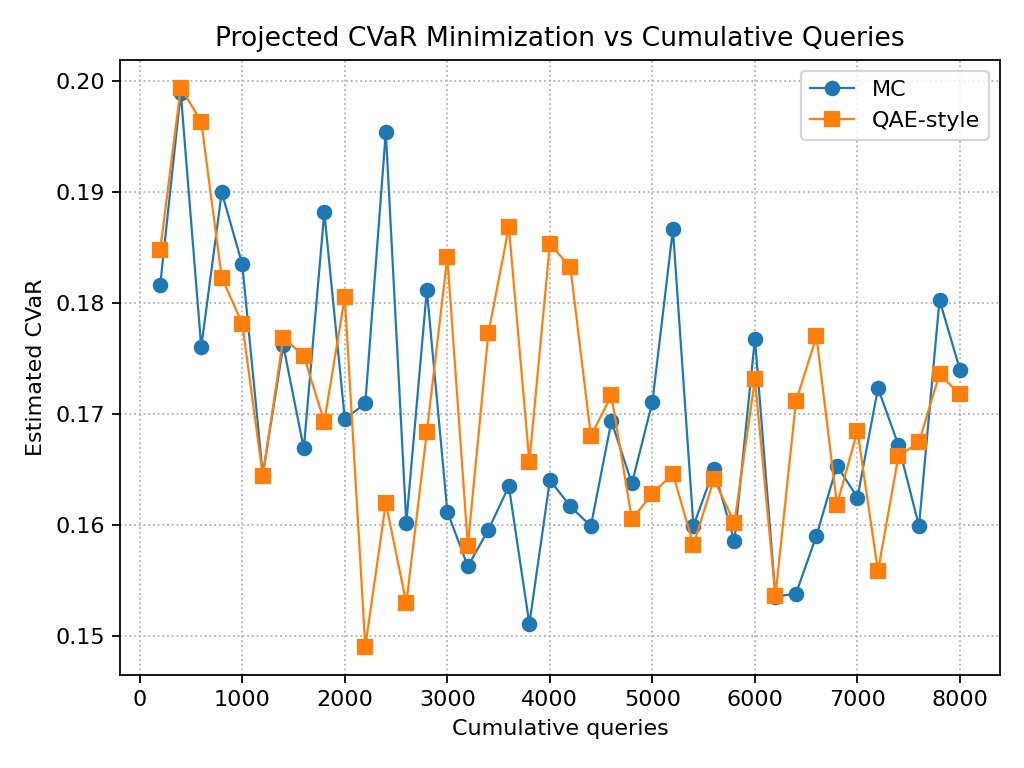}
    \caption{Projected CVaR minimization plotted against cumulative queries. Under matched per-iteration query budgets the trajectories coincide; the resource advantage of QAE-style estimation appears at the gradient-oracle level (Figure~\ref{fig:grad_error}, Table~\ref{tab:grad_error}).}
    \label{fig:opt_queries}
\end{figure}

We next assess the downstream impact on optimization. Using projected stochastic subgradient descent with a fixed per-iteration budget, we track the estimated CVaR across iterations. Figures~\ref{fig:opt_iters} and~\ref{fig:opt_queries} visualize these results. When plotted against iterations, both methods show comparable improvements in CVaR (Figure~\ref{fig:opt_iters}), which is the expected behavior at matched per-iteration budgets: both oracles deliver gradient estimates of sufficient accuracy that the dominant source of error is the $O(1/\sqrt{T})$ SGD term in Proposition~\ref{prop:sgd-convergence}, not the oracle bias $\epsilon$. The trajectories track each other within statistical noise (mean CVaR over the trajectory: $0.1695$ for MC versus $0.1708$ for QAE-style; see Table~\ref{tab:opt_traj}).

The query-efficiency advantage of QAE only manifests when budgets are varied across the two methods. Because we matched the per-iteration query count exactly in this experiment, Figure~\ref{fig:opt_queries} replots the same trajectories against cumulative queries and shows the two curves essentially superimposed. To translate the gradient-level scaling in Section~\ref{subsec:grad-accuracy} into an optimization-level resource saving, one would equate the MC error $O(1/\sqrt{N})$ to the QAE-style error $O(1/M)$, giving $M=\Theta(\sqrt{N})$: the QAE-style oracle attains the same per-iteration gradient accuracy with a quadratically smaller query budget, so the total query count to reach a given CVaR trajectory scales as $\sqrt{N}$ times fewer queries. The present optimization experiment is therefore best read as a sanity check that the QAE-style oracle drives the SGD loop to the same CVaR minimum as MC, with the resource gain to be demonstrated separately at the gradient-estimation level.

\subsection{Discussion}

Taken together, the numerical evidence and figures are consistent with the theoretical predictions of the paper. The gradient estimation study provides clear empirical support for the quadratic improvement in error scaling afforded by amplitude estimation: at matched budgets, the QAE-style estimator achieves more than an order of magnitude lower error, and the $N=M^2$ overlay verifies the predicted slopes. The optimization experiments confirm that this improved oracle accuracy translates without loss to the projected SGD dynamics, with both methods reaching the same CVaR minimum. Combined, these results substantiate the per-iteration $O(d/\epsilon)$ versus $O(d/\epsilon^2)$ gap of Proposition~\ref{prop:qae-gradient}, and motivate the overall $\widetilde O(d/\epsilon^3)$ versus $\widetilde O(d/\epsilon^4)$ complexity of Proposition~\ref{prop:sgd-convergence} for risk-sensitive portfolio optimization in quantum finance.

\section{Future Work}

Our results open several promising avenues for further investigation. Below we highlight key directions for strengthening, generalizing, and applying the quantum subgradient methodology in practical and theoretical settings.

\subsection{Noise-robustness and fault tolerance}  
In this work we assumed an ideal (noiseless) amplitude-estimation (AE) oracle. A natural next step is to analyze the behavior of our quantum subgradient oracle under realistic noise models (e.g.\ depolarizing noise, measurement error, finite coherence time) and to derive robust bounds on the bias and variance propagation. Recent quantum convex optimization schemes with noisy evaluation oracles (e.g.\ \cite{Augustino2025}) may offer useful techniques for this extension. Or even apply these techniques in real hardware~\cite{Agliardi2025SamplingCVaRVQA}.

\subsection{Accelerated and mirror-space methods}  
We used a straightforward stochastic subgradient descent approach. It would be fruitful to extend our framework to accelerated methods (e.g.\ Nesterov acceleration) or mirror-descent and dual-averaging in non-Euclidean geometries. The recent work by Augustino et al.\ on dimension-independent quantum gradient methods suggests that such advanced algorithms may yield improved worst-case complexity~\cite{Augustino2025}.

\subsection{Beyond linear losses: nonlinear payoffs and derivative portfolios}  
Our analysis assumes a portfolio with linear returns (i.e.\ inner product \(w^\top R\)). Extending our quantum subgradient oracle to nonlinear loss functions, such as option payoff portfolios or other nonlinear financial instruments, is a compelling challenge. This would require developing quantum circuits for conditional expectations over nonlinear mappings and controlling associated bias.

\subsection{Hybrid heuristics and variational hybrids}  
Given that near-term quantum devices may not support deep AE circuits, one could explore hybrid methods combining our oracle with variational or sampling-based subroutines. For instance, integrating subgradient estimates into VQA / CVaR heuristics (like in~\cite{Barkoutsos2020}) or using local classical post-processing could yield practical performance on NISQ hardware.

\subsection{Resource trade-off and empirical scaling on quantum hardware}  
While we provide a resource estimate in Appendix~\ref{app:resources}, a more detailed study of circuit depth, qubit routing overhead, measurement repetition overhead, and error mitigation trade-offs for varying problem sizes (dimension \(d\), target error \(\varepsilon\)) would help bridge theory and practice. Empirical tests on intermediate-scale quantum devices would validate the scaling constants.

\subsection{Lower bounds and optimality regimes}  
We have shown an upper bound of \( \widetilde O(d / \varepsilon) \) per subgradient estimate, but a matching lower bound specifically for CVaR subgradient estimation remains open. A lower bound tailored to the conditional-expectation structure (akin to ITCS-style bounds for nonsmooth convex optimization) would clarify whether further quantum improvements are possible.

\subsection{Multiple risk measures and multi-objective optimization}  
Finally, extending the quantum subgradient framework to other coherent risk measures (e.g.\ entropic risk, spectral risk measures) or to \emph{multi-objective optimization} (e.g.\ minimizing CVaR under return constraints) would broaden the applicability to more realistic financial decision problems.

Overall, these directions together point toward a richer theory of quantum risk optimization and bring us closer to practical quantum-enhanced methods for financial risk management.

\bibliographystyle{unsrt}  
\bibliography{references}  

\appendix

\section{Preliminaries}
\label{sec:preliminaries}

We consider a portfolio of $d$ assets with weight vector $w \in \mathcal{W} \subseteq \mathbb{R}^d$, 
where $\mathcal{W}$ denotes the feasible set (e.g., the probability simplex for long-only portfolios).
Let the random return vector be $r \sim \mathcal{D}$, where $\mathcal{D}$ is a fixed but unknown distribution
estimated from historical data or a factor model. 
The associated portfolio \emph{loss} is
\[
L(w) = - w^\top r.
\]

\paragraph{Expectation operator.}
Throughout, $\mathbb{E}[\cdot]$ denotes expectation with respect to the return distribution $r \sim \mathcal{D}$, 
equivalently over the induced distribution of $L(w)$.
When we analyze stochastic algorithms (e.g.\ projected subgradient descent), 
$\mathbb{E}[\cdot]$ will additionally encompass randomness due to 
the algorithm itself and due to oracle estimation noise.

\paragraph{Value-at-Risk (VaR).}
For confidence level $\alpha \in (0,1)$, the \emph{Value-at-Risk} is the $\alpha$-quantile of the loss distribution:
\[
\mathrm{VaR}_\alpha(w) = \inf \{ z \in \mathbb{R} : \Pr[L(w) \leq z] \geq \alpha \}.
\]

\paragraph{Conditional Value-at-Risk (CVaR).}
The \emph{Conditional Value-at-Risk} (also known as expected shortfall) is the expected loss 
beyond the VaR threshold~\cite{Rockafellar2000,Rockafellar2002}:
\[
\mathrm{CVaR}_\alpha(w) = \mathbb{E}\!\left[ L(w) \,\middle|\, L(w) \geq \mathrm{VaR}_\alpha(w) \right].
\]
Equivalently, CVaR admits the optimization-based representation
\[
\mathrm{CVaR}_\alpha(w) 
= \min_{z \in \mathbb{R}} \Bigg\{ z + \frac{1}{1-\alpha}\,
  \mathbb{E}\big[ (L(w)-z)_+ \big] \Bigg\},
\]
where $(x)_+ = \max\{x,0\}$.
This convex formulation is the basis of the Rockafellar--Uryasev approach to CVaR minimization.

\paragraph{Subgradients of CVaR.}
A subgradient of $\mathrm{CVaR}_\alpha$ with respect to $w$ is given by
\[
g(w) = \mathbb{E}\!\left[ \nabla_w L(w) \,\middle|\, L(w) \geq \mathrm{VaR}_\alpha(w) \right],
\]
as shown in~\cite{Rockafellar2000,Rockafellar2002}.
For linear losses $L(w)=-w^\top r$, the gradient is $\nabla_w L(w) = -r$.

\paragraph{Computational bottleneck.}
Both $\mathrm{CVaR}_\alpha(w)$ and its subgradient $g(w)$ are expectations 
over the \emph{tail event} $\{L(w) \geq \mathrm{VaR}_\alpha(w)\}$, 
which are typically estimated by Monte Carlo methods.
Classical sampling requires $O(1/\epsilon^2)$ scenarios to achieve 
an $\epsilon$-accurate estimate of such conditional expectations.
Quantum Amplitude Estimation (QAE)~\cite{Brassard2002,Montanaro2015,Suzuki2020} 
reduces this to $O(1/\epsilon)$ queries, motivating our study of quantum CVaR subgradient oracles.

\section{Proofs and Conditions of the Main Propositions}
\label{proofs}

\subsection{Proof of Proposition~\ref{prop:threshold-bias} (Bias from VaR threshold error)}
\label{app:bias}

Let $z_\alpha = \mathrm{VaR}_\alpha(w)$ and $\tilde{z}$ be the approximate threshold.
Define
\[
\mu(z) := \mathbb{E}\big[\nabla_w L(w)\,\mathbf{1}\{L(w)\geq z\}\big], 
\quad
p(z) := \Pr[L(w)\geq z].
\]
Then
\[
g(w) = \frac{\mu(z_\alpha)}{p(z_\alpha)}, \qquad
\tilde{g}(w) = \frac{\mu(\tilde{z})}{p(\tilde{z})}.
\]

\paragraph{Step 1: Bound numerator difference.}
Suppose $\|\nabla_w L(w)\|_2 \leq G$ almost surely and the density of $L(w)$ exists and is bounded by $M$. Then
\begin{align*}
\|\mu(\tilde{z}) - \mu(z_\alpha)\|_2
&= \left\|\mathbb{E}[\nabla_w L(w)\,(\mathbf{1}\{L(w)\geq \tilde{z}\} - \mathbf{1}\{L(w)\geq z_\alpha\})]\right\|_2 \\
&\leq G \Pr\big(z_\alpha \leq L(w) < \tilde{z}\big) \\
&\leq G M |\tilde{z} - z_\alpha|.
\end{align*}

\paragraph{Step 2: Bound denominator difference.}
Similarly,
\[
|p(\tilde{z}) - p(z_\alpha)| \leq M |\tilde{z} - z_\alpha|.
\]

\paragraph{Step 3: Ratio perturbation.}
We use the inequality
\[
\left\|\frac{a}{b} - \frac{c}{d}\right\|_2 \leq \frac{\|a-c\|_2}{|b|} + \frac{\|c\|_2}{|b||d|}\,|b-d|.
\]
With $a=\mu(\tilde{z}), c=\mu(z_\alpha), b=p(\tilde{z}), d=p(z_\alpha)$, both differences are $O(|\tilde{z}-z_\alpha|)$. Therefore
\[
\|\tilde{g}(w) - g(w)\|_2 = O(|\tilde{z}-z_\alpha|).
\]

\paragraph{Conclusion.} Setting $\delta = |\tilde{z} - z_\alpha|$ yields the claim.
\qed

\subsection{Proof of Proposition~\ref{prop:qae-gradient} (Quantum query complexity)}
\label{app:query}

We recall that $g(w) = \mu(z)/p(z)$ with $\mu(z)\in\mathbb{R}^d$.

\paragraph{Step 1: Estimation via QAE.}
For each coordinate $j=1,\ldots,d$, define the bounded random variable
\[
X_j = \big(\nabla_w L(w)\big)_j \cdot \mathbf{1}\{L(w)\geq z\}, \qquad |X_j|\leq G.
\]
QAE estimates $\mathbb{E}[X_j]$ to additive error $\epsilon_j$ using $O(1/\epsilon_j)$ queries~\cite{Suzuki2020}. Similarly, $p(z)$ is estimated to error $\epsilon_p$ with $O(1/\epsilon_p)$ queries.

\paragraph{Step 2: Vector accuracy.}
To ensure $\|\hat{\mu}-\mu\|_2 \leq \epsilon/2$, it suffices to set $\epsilon_j = \epsilon/\sqrt{d}$ for each coordinate. This requires $O(\sqrt{d}/\epsilon)$ queries per coordinate, i.e.\ $O(d/\epsilon)$ in total.

\paragraph{Step 3: Error propagation.}
We write
\[
\|\tilde{g}(w)-g(w)\|_2 \leq \frac{\|\hat{\mu}-\mu\|_2}{p(z)} + \frac{\|\mu\|_2}{p(z)^2}|\hat{p}-p(z)|.
\]
Both terms can be bounded by $O(\epsilon)$ using QAE with $\epsilon$ accuracy on numerator and denominator.

\paragraph{Step 4: Success probability.}
Amplifying confidence via repetition and median-of-means increases complexity only by $\log(1/\eta)$.

\paragraph{Conclusion.}
The total query complexity is
\[
T = O\!\left(\frac{d}{\epsilon}\log\frac{1}{\eta}\right).
\]
In contrast, Monte Carlo requires $O(d/\epsilon^2)$ samples.
\qed

\subsection{Proof of Proposition~\ref{prop:sgd-convergence} (Projected SGD convergence)}
\label{app:sgd}

Let $f(w)=\mathrm{CVaR}_\alpha(w)$, convex with bounded subgradients.

\paragraph{Step 1: Noisy SGD bound.}
Classical results for projected stochastic subgradient descent with inexact gradients (e.g.\ \cite{Nemirovski2009}) show that if
\[
\mathbb{E}\big[\|\tilde{g}(w)-g(w)\|_2\big] \leq \epsilon,
\]
then with step-size $\eta_t = O(1/\sqrt{t})$,
\[
\min_{t\leq T} \mathbb{E}[f(w_t)-f(w^\star)] \leq O\!\left(\frac{1}{\sqrt{T}} + \epsilon\right).
\]

\paragraph{Step 2: Oracle cost.}
To achieve error $\epsilon$, we need $T=O(1/\epsilon^2)$ iterations. Each iteration requires one $\epsilon$-accurate gradient oracle, costing $O(d/\epsilon)$ queries by Proposition~\ref{prop:qae-gradient}.

\paragraph{Step 3: Total complexity.}
Thus total queries are
\[
O\!\left(\frac{d}{\epsilon}\cdot \frac{1}{\epsilon^2}\right) = O\!\left(\frac{d}{\epsilon^3}\right).
\]

\paragraph{Step 4: Classical comparison.}
Monte Carlo requires $O(d/\epsilon^2)$ samples per gradient, leading to $O(d/\epsilon^4)$ overall. Therefore, the quantum method yields a near-quadratic improvement.

\qed

\section{Resource Analysis: Physical Qubit Requirements}
\label{app:resources}

A central question for practical deployment of QAE-based CVaR optimization is how many physical qubits would be required on near- or mid-term hardware to realize the algorithm at useful scales. While our numerical experiments emulate the quadratic query advantage of amplitude estimation, mapping this into physical resources requires careful consideration of logical qubits, error correction, and overhead.

\subsection{Logical Qubit Estimates}

The core task in QAE-style CVaR estimation is preparing a distributional oracle that encodes portfolio losses into amplitudes, and then performing phase estimation to extract probabilities. For a portfolio with $d$ assets and budget discretization into $B$ bins, the state preparation register requires approximately
\[
n_{\text{data}} \approx \lceil \log_2 (d \cdot B) \rceil
\]
logical qubits. Additional ancillae are required for arithmetic (summing weighted returns, comparisons against VaR thresholds) and the amplitude estimation circuit itself. In total, a minimal logical requirement is in the range of
\[
n_{\text{logical}} \sim n_{\text{data}} + \mathcal{O}(\log(1/\epsilon)),
\]
where $\epsilon$ is the target precision of the CVaR estimate. For typical experimental settings ($d=10$, $B=2^{10}$ bins, $\epsilon \approx 10^{-2}$), this corresponds to roughly $n_{\text{logical}} \approx 20\!-\!30$ logical qubits.

\subsection{Error Correction Overheads}

Current quantum hardware is noisy, and running QAE circuits at depth requires fault-tolerant encoding. Surface code error correction is the leading candidate, with physical-to-logical overhead scaling approximately as
\[
n_{\text{physical}} \approx \alpha \cdot n_{\text{logical}} \cdot d_{\text{code}}^2,
\]
where $d_{\text{code}}$ is the code distance required to suppress logical errors to acceptable levels and $\alpha$ is a constant accounting for layout. For error rates $p \approx 10^{-3}$ and target logical failure probabilities $\approx 10^{-9}$, a distance of $d_{\text{code}} \sim 30$ is typical, leading to overhead factors in the range of $10^3$ physical qubits per logical qubit.

Thus, the physical resource count becomes
\[
n_{\text{physical}} \sim 20\!-\!30 \times 10^3 \approx 2\!-\!3 \times 10^4
\]
physical qubits to execute CVaR estimation with portfolio dimension $d=10$ at precision $\epsilon \sim 10^{-2}$.

\subsection{Scalability and Implications}

The above analysis demonstrates two important points:

\begin{enumerate}
    \item The logical qubit requirements of QAE-style CVaR estimation scale only logarithmically with portfolio discretization and polynomially with target precision, making the algorithm theoretically scalable.
    \item However, current fault-tolerant overheads inflate the physical qubit count into the tens of thousands even for modest instances. This places near-term implementation out of reach but provides a concrete target for hardware roadmaps.
\end{enumerate}

\subsection{Perspective}

While tens of thousands of physical qubits are beyond today's devices, several technology trends can reduce this barrier. Improved error rates would reduce code distances, hardware-efficient encodings could shrink arithmetic costs, and hybrid quantum--classical methods may amortize some of the heavy lifting. Our analysis therefore frames the resource requirements not as a barrier but as a benchmark: once fault-tolerant devices with $\sim 10^5$ physical qubits become available, QAE-based CVaR optimization will be a realistic candidate application of quantum computing to financial risk management.

\section{Acknowledgments and Code Availability}
The authors declare no competing interests.\\

The code used for the data processing and presentation is maintained and available at \url{https://github.com/BillSkarlatos/Quantum-CVaR} for the reader to clone and reproduce those results.

\section{Plots and Tables}

\begin{table}[h]
\centering
\begin{tabular}{|c|c|c|}
\hline
Method & Budget & Gradient $\ell_2$ Error \\
\hline
MC & 100 & 0.31862 \\
\hline
MC & 215 & 0.14953 \\
\hline
MC & 464 & 0.09620 \\
\hline
MC & 1000 & 0.09827 \\
\hline
MC & 2154 & 0.05187 \\
\hline
MC & 4641 & 0.03030 \\
\hline
MC & 10000 & 0.01753 \\
\hline
QAE-style & 10 & 0.36073 \\
\hline
QAE-style & 21 & 0.19755 \\
\hline
QAE-style & 46 & 0.07822 \\
\hline
QAE-style & 100 & 0.01017 \\
\hline
QAE-style & 215 & 0.01352 \\
\hline
QAE-style & 464 & 0.00552 \\
\hline
QAE-style & 1000 & 0.00446 \\
\hline
\end{tabular}
\caption{Numerical values of CVaR gradient $\ell_2$ errors across budgets. At matched budgets ($N=M$), the QAE-style estimator achieves more than an order of magnitude lower error than MC (e.g.\ $0.0102$ vs.\ $0.319$ at budget $100$; $0.00446$ vs.\ $0.0983$ at budget $1000$), and the per-budget slope is consistent with the predicted $O(1/M)$ versus $O(1/\sqrt{N})$ scaling. Cross-method averages over heterogeneous budgets are not reported here, as the two sweeps cover different budget ranges and such averages do not characterize asymptotic scaling.}
\label{tab:grad_error}
\end{table}

\begin{table}[h]
\centering
\begin{tabular}{|c|c|c|c|c|}
\hline
Iter & CVaR (MC) & Queries (MC) & CVaR (QAE-style) & Queries (QAE-style) \\
\hline
1 & 0.18163 & 200 & 0.18479 & 200 \\
\hline
2 & 0.19895 & 400 & 0.19977 & 400 \\
\hline
3 & 0.17574 & 600 & 0.18706 & 600 \\
\hline
4 & 0.18340 & 800 & 0.17777 & 800 \\
\hline
5 & 0.17680 & 1000 & 0.16439 & 1000 \\
\hline
6 & 0.16676 & 1200 & 0.17479 & 1200 \\
\hline
7 & 0.16662 & 1400 & 0.17613 & 1400 \\
\hline
8 & 0.18857 & 1600 & 0.16792 & 1600 \\
\hline
9 & 0.17020 & 1800 & 0.16882 & 1800 \\
\hline
10 & 0.17107 & 2000 & 0.14880 & 2000 \\
\hline
11 & 0.19388 & 2200 & 0.16243 & 2200 \\
\hline
12 & 0.16013 & 2400 & 0.15352 & 2400 \\
\hline
13 & 0.18104 & 2600 & 0.16767 & 2600 \\
\hline
14 & 0.15675 & 2800 & 0.18449 & 2800 \\
\hline
15 & 0.15886 & 3000 & 0.16762 & 3000 \\
\hline
16 & 0.15938 & 3200 & 0.16686 & 3200 \\
\hline
17 & 0.15107 & 3400 & 0.18344 & 3400 \\
\hline
18 & 0.17862 & 3600 & 0.16921 & 3600 \\
\hline
19 & 0.16431 & 3800 & 0.18718 & 3800 \\
\hline
20 & 0.17179 & 4000 & 0.16546 & 4000 \\
\hline
21 & 0.17123 & 4200 & 0.18324 & 4200 \\
\hline
22 & 0.15988 & 4400 & 0.16800 & 4400 \\
\hline
23 & 0.16720 & 4600 & 0.17111 & 4600 \\
\hline
24 & 0.17107 & 4800 & 0.16006 & 4800 \\
\hline
25 & 0.18652 & 5000 & 0.16483 & 5000 \\
\hline
26 & 0.17129 & 5200 & 0.16472 & 5200 \\
\hline
27 & 0.15823 & 5400 & 0.15829 & 5400 \\
\hline
28 & 0.16477 & 5600 & 0.16444 & 5600 \\
\hline
29 & 0.15822 & 5800 & 0.16004 & 5800 \\
\hline
30 & 0.17661 & 6000 & 0.17347 & 6000 \\
\hline
31 & 0.15324 & 6200 & 0.15409 & 6200 \\
\hline
32 & 0.15905 & 6400 & 0.17115 & 6400 \\
\hline
33 & 0.15891 & 6600 & 0.17693 & 6600 \\
\hline
34 & 0.16539 & 6800 & 0.16223 & 6800 \\
\hline
35 & 0.17138 & 7000 & 0.16788 & 7000 \\
\hline
36 & 0.15114 & 7200 & 0.15647 & 7200 \\
\hline
37 & 0.16005 & 7400 & 0.16536 & 7400 \\
\hline
38 & 0.18032 & 7600 & 0.17557 & 7600 \\
\hline
39 & 0.17368 & 7800 & 0.17674 & 7800 \\
\hline
40 & 0.17400 & 8000 & 0.17252 & 8000 \\
\hline
\textbf{Average} & \textbf{0.16947} & \textbf{-} & \textbf{0.17076} & \textbf{-} \\
\hline
\end{tabular}
\caption{Optimization trajectories with CVaR and cumulative queries reported at each iteration. With matched per-iteration query budgets, the two methods reach nearly identical mean trajectory CVaR ($0.1695$ for MC vs.\ $0.1708$ for QAE-style), confirming that the QAE-style oracle drives the SGD loop to the same optimum as MC. The query-efficiency advantage of QAE manifests at the gradient-estimation level (Table~\ref{tab:grad_error}).}
\label{tab:opt_traj}
\end{table}

\end{document}